\documentstyle[multicol,aps,prl]{revtex}
\begin{document}

\title{Wide ultrarelativistic plasma beam -- magnetic barrier collision
and astrophysical applications}

\author{V.V. Usov and M.V. Smolsky}

\address{Department of Condensed Matter Physics, Weizmann Institute,
Rehovot 76100, Israel}

\maketitle

\begin{abstract}
  
  The interaction between a wide ultrarelativistic fully-ionized
plasma beam and a magnetic barrier is studied numerically.
It is assumed that the plasma beam is initially homogeneous and impacts 
with the Lorentz factor $\Gamma_0\gg 1$ on the barrier. The magnetic
field of the barrier ${\bf B}_0$ is uniform and transverse to the beam 
velocity. When the energy densities of the beam and the magnetic field are 
comparable, $\alpha = 8\pi n_0m_pc^2(\Gamma_0-1)/B^2_0\sim 1$, the process
of the beam -- barrier interaction is strongly nonstationary, and the 
density of reversed protons is modulated in space by a factor of 10 or so.
The modulation of reversed protons decreases with decrease of $\alpha$.
The beam is found to penetrate deep into the barrier provided 
that $\alpha > \alpha _{\rm cr}$, where $\alpha _{\rm cr}$ is 
about 0.4. The speed of such a penetration is subrelativistic
and depends on $\alpha$. 
Strong electric fields are generated near the front 
of the barrier, and electrons are accelerated in these fields up to the
mean energy of protons, i.e. up to $\sim m_pc^2\Gamma_0$. 
The synchrotron radiation of high-energy  electrons from
the front vicinity is calculated. Stationary solutions for the beam --
barrier collision are considered. It is shown that such a solution
may be only at $\alpha \lesssim 0.2 - 0.5$ depending on the boundary 
conditions for the electric field in the region of the
beam -- barrier interaction. Some astrophysical
applications of these results are briefly discussed.

\end{abstract}


\begin{multicols}{2}

\section{Introduction}

There is now compelling evidence that plasma is ejected from many
astronomical objects and flows away at relativistic speeds. The
Lorentz factor, $\Gamma_0$, of such plasma wind is a few $\times
(1-10)$ for the jets associated with active galactic nuclei
\cite{Cawthorne91} and $\sim 10^2-10^3$ or even more for the $\gamma$-ray
bursters \cite{Baring93}. A strong magnetic field may be in the
outflowing gas \cite{Narayan92,Usov92,Usov94,TD}.  Relativistic
magnetized winds can interact with an external medium (e.g., an ordinary 
interstellar medium). It was pointed out in Ref. \cite{Meszaros93}
that such an interaction may be responsible for radiation of both
X-rays and $\gamma$-rays from the $\gamma$-ray bursters.
For consideration of the interaction between a
relativistic magnetized wind and an external medium, it is convenient
to switch to the wind frame. In this frame, the problem of the magnetized
wind -- external medium interaction is identical to the problem of
collision between a wide relativistic beam of cold plasma and a region
with a strong magnetic field which is called a magnetic barrier.

The problem of the interaction between plasma beams and  magnetic 
barriers was attacked in many experiments and theoretical papers (see Ref. 
\cite{Gordeev94} and references therein). However, all known experimental
studies of the beam -- magnetic barrier interaction are irrelevant to
our problem. This is because the plasma beams produced by the laboratory
equipment are either non-relativistic or very narrow and the cross-beam sizes 
are much smaller than the gyroradius of the beam protons in the field of
the barrier. In the theoretical studies,
it is usually used the assumption of Rosenbluth \cite{Ros63} that the
charge separation is small. However, this assumption is valid provided 
that the inequality $(v_0/c)^2\ll m_e/m_p\simeq 5.4\times 10^{-4}$ 
holds true, where $v_0$ is the beam velocity,
$m_e$ is the electron mass and $m_p$ is the proton mass.
For relativistic beams, $v_0 = c\sqrt{1-(1/\Gamma_0)^{-2}} 
\simeq c$, the assumption of Rosenbluth proves inadequate, 
and the charge separation is very important for the dynamics of 
particles near the barrier front \cite{Peter79,Smolsky95}.
Below, the interaction of a wide ultrarelativistic plasma beam 
with a magnetic barrier is considered numerically (for preliminary 
results and some astrophysical applications, see Ref. \cite{Smolsky95}).

\section{Formulation of the problem and basic equations}

The situation to be discussed is the following. At the initial moment,
$t=0$, the ultrarelativistic (Lorentz factor $\Gamma_0\gg 1$) neutral 
beam of protons and electrons (number densities $n_p=n_e\equiv n_0$) runs
along the $x$ axis into the magnetic barrier which is 
the half-space $x>0$ with an external magnetic field
${\bf B}_0=B_0{\bf \hat e}_z\Theta[x]$, where $n_0$,
$B_0$ are constants and $\Theta [x]$ is the step function equal to
unity for $x>0$ and to zero for $x<0$. The beam is infinite in the $y$
-- $z$ dimensions and semi-infinite in the $x$ dimension. Our goal is to 
construct a 1${1\over 2}$D time-dependent solution for the problem, i.e. to
find induced electromagnetic fields (${\bf E}=E_x{\bf \hat
  e}_x+E_y{\bf \hat e}_y$; ${\bf B}=B{\bf \hat e}_z$) and 
motion of the beam particles
in the $x$ -- $y$ plane. The field structure and 
the beam particle motion are to be
treated self-consistently. All the quantities are assumed to be
dependent on $t$ and $x$ only.

As noted earlier, the strength of the magnetic fields in the 
astrophysical winds may be very high, especially in the winds 
outflowing from the $\gamma$-ray bursters \cite{Usov92,Usov94}.
High-energy electrons generate synchrotron radiation in these fields,   
and the radiation damping force which acts on the radiating electrons
has to be taken into account. Since $m_p\gg m_e$, both the synchrotron
radiation of protons and the radiation damping force which acts on the 
protons are very small (e.g., Ref. \cite{Landau71}) and may be neglected.

The following set of equations can be used to describe the process
of the beam -- barrier collision \cite{Landau71}:

\begin{eqnarray}   
\frac{\partial E_x}{\partial x} &=& 4\pi \rho \,, \label{dExdx} \\
\frac{\partial E_y}{\partial x} &=&
  -\frac 1c \frac{\partial B}{\partial t}\,, \label{dEydx} \\
\frac{\partial B}{\partial x} &=& -\frac 1c\frac{\partial E_y}
{\partial t}-\frac{4\pi }cj_y\,, \label{dBdx} \\
mc\frac{du^i}{ds} &=& \frac ec F^{ik}u_k+\frac{2e^4}{3m^2c^5}
(F_{kl}u^l)(F^{km}u_m)u^l\,, \label{Motion}
\end{eqnarray}

\noindent 
where $\rho$ and ${\bf j}$ are the densities of charges and currents
respectively, $e$ is the charge of particles, $m$ is the mass of 
particles, $m=\{m_e,\,m_p\}$, $s$ is the interval,
$u^i=\{\Gamma,({\bf v}/c)\Gamma\}$ is the four-velocity, $\Gamma$ is
the Lorentz factor of particles, and $F^{ik}$ is the electromagnetic field
tensor. No simplifying
assumptions besides geometrical ones are exploited when rewriting
Maxwell equations (\ref{dExdx}) -- (\ref{dBdx}). The second term in 
the right-hand side of the relativistic equation of particle motion 
(\ref{Motion}) is the radiation damping four-force. Following Ref.
\cite{Smolsky95}, only the term of the highest power in $\Gamma$ 
is left in the damping force. This is valid for ultrarelativistic motion
of particles, $\Gamma \gg 1$ (e.g., Ref. \cite{Landau71}). 

The four-velocity $u^i$ from Eq. (\ref{Motion}) is related to 
the charge and current densities which appear in the Maxwell equations
(\ref{dExdx}) -- (\ref{dBdx}): 

\begin{equation}
j^i=\{c\rho,j_x,j_y,0\}=\sum
ce\Gamma^{-1}u^i\,, 
\end{equation}

\noindent
where $\sum$ means the sum over the beam particles
per unitary volume.

The energy losses, $I$, of electrons because of their radiation in 
the electric and magnetic fields are computed
using the following equation \cite{Landau71}:
 
\begin{equation}
I=\frac{2e^4}{3m_e^2c^3}\left\{ \left( {\bf E} +\frac 1c{\bf v
\times B}\right) ^2-\frac 1{c^2}\left( {\bf E\cdot v}\right)^2 \right\}
\Gamma ^2\,.
\end{equation}

To evaluate the spectrum of the radiation of electrons,
we have used the following expression for the spectral intensity
of synchrotron radiation \cite{RL79}:
 
\begin{equation}
I_\nu
=\frac {\sqrt{3}e^3B }{m_ec^2} \frac \nu {\nu_c}
\int_{\nu /\nu_c}^{\infty}K_{5/3} (\eta )d\eta \,,\label{Inu}
\end{equation}
 
\noindent where
$K_{5/3}$ is the modified Bessel functions of 5/3 order,
 
\begin{equation}
\nu_c={3eB\Gamma ^2\over 4\pi m_ec}\,.\label{nuc}
\end{equation}
 
In Eqs. (\ref{Inu}) and (\ref{nuc}), it is taken into account that in our
case the velocities of particles are perpendicular to the magnetic
field, ${\bf v}\perp {\bf B}$.

The initial beam number density is taken from the equation:

\begin{equation}
n_0m_pc^2(\Gamma _0-1)=\alpha \frac{B_0^2}{8\pi}\,, \label{DefAlpha}
\end{equation}

\noindent where $\alpha$ is the dimensionless parameter. In this work
we study mainly the case $\alpha\sim 1$ when the plasma flow pressure
is comparable with the magnetic field pressure of the barrier. 
In different runs, $\alpha$ ranges from 0.2 to 4.

To integrate the set of equations (\ref{dExdx}) -- (\ref{DefAlpha}), we
used a macro-particle approximation in which all particles of the beam
are subdivided into a large number, $\sim 10^4$, of packets. The
particles within a packet are bundled together forming a large
``macro-particle'' which is infinitively small along the axis $x$ and
infinitively large in all directions of the $y,z$ plane. Each packet
contains either protons or electrons (for details of the numerical
method, see Ref. \cite{Smolsky95}). 

The examined space-time domain is the
following:

\begin{equation}
x_{\min}<x<x_{\max} \,\,\,\,\,{\rm and}\,\,\,\,\, 0<t<t_{\max}\,, 
\end{equation}

\noindent
where $x_{\min }=-$
a few $\times (1-10)(c/\omega_{Bp})$, $x_{\max}=$ a few $\times (1-10)
(c/\omega_{Bp})$, $t_{\max}=$ a few $\times (1-10)
T_p$, $T_p=2\pi /\omega_{Bp}$
is the proton gyroperiod, and  $\omega_{Bp}=eB_0/(m_pc\Gamma_0)$ and
$c/\omega_{Bp}=m_pc^2\Gamma_0/(eB_0)$ are the proton
gyrofrequency and gyroradius, respectively. The time step
of calculations is a few $\times 10^{-6}T_p$, depending on $\alpha$.

The boundary condition for $E_x$ is
$E_x\left| _{x=x_{\max}}\right.=0$, as $x_{\max}$ is chosen so that no
beam particles penetrate deeper than $x_{\max}$ in the observed
interval of time. Electromagnetic waves generated due to charge flow
are allowed to escape freely from the system that is used as
boundary conditions for $B$ and $E_y$ in Eqs. (\ref{dEydx}) and
(\ref{dBdx}). 

\section{Results of numerical simulations and scaling}

The relativistic cold plasma -- magnetic barrier collision is 
characterized by the following parameters: $B_0$, $\Gamma_0$ and 
$\alpha$. The input parameters of our simulations are given in Table  
\ref{Params}. The values of $B_0$ and $\Gamma_0$ are chosen to be relevant 
to cosmological $\gamma$-ray bursters 
\cite{Narayan92,Usov92,Usov94,Meszaros93,Smolsky95}.

\subsection{Particle dynamics and penetration of the beam particles
into the barrier}

The density of protons which move towards the 
magnetic barrier, $v_x>0$, is almost
unperturbed untill the distance to the barrier is smaller than $\sim
c/\omega_{Bp}$. Reversed protons which move away from 
the magnetic barrier, $v_x<0$, are bunched 
in the process of the beam -- barrier interaction (Fig. \ref{ProtDens}). 
The modulation of the density of reversed protons increases with increase
of $\alpha$. Namely, the ratio of the maximum to minimum densities of
reversed protons is $\sim 2$ at $\alpha \simeq 0.2$ and $\sim
10$ at $\alpha \simeq 1$.  A typical length of such a modulation is
roughly the proton gyroradius $c/\omega_{Bp}$. Similar phenomena was
observed also in numerical simulations of collisionless shock waves 
near the shock front (e.g., Ref. \cite{Hoshino92}). 

At $\alpha\sim 1$, the mean Lorentz factor of reversed protons
that far enough from the barrier, $x<-(c/\omega_{Bp})$, where the
process of strong interaction between particles and fields is 
more or less over, is
$\langle \Gamma_p^{\rm{out}}\rangle \simeq (0.5\pm 0.1)\Gamma_0$,
i.e. about half of the initial kinetic energy of protons is lost in
the process of their collision with the barrier (see Table 
\ref{Results}). Figure \ref{ProtSpect} shows the 
energy distribution for reversed protons. 

The density of electrons which move away from the
barrier is well correlated with the density of reversed protons; the
correlation coefficient is $r\simeq 0.8$, where
$r=\sigma_{pe}/\sqrt{\sigma_{pp}\sigma_{ee}}$, and

\begin{equation}
\sigma _{ij}=\int_{x_{\min}}^{x_{\max}}
(n_i-\overline{n_i})(n_j-\overline{n_j})dx\,,\,\,\,\, 
\{i,j\}=\{e,p\}\,.
\end{equation}

\noindent These electrons screen the electrostatic field, $E_x$, of
proton bunches mostly, but not completely.

At $\alpha > \alpha _{\rm cr}$, $\alpha_{\rm cr}\simeq 0.4$, it is 
observed that the length of the beam particle
penetration into the barrier increases in time. Figure \ref{Penetr}
shows the $x$ coordinate, $x_{\rm pen}$, of the most deeply
penetrated proton as a function of time $t$. 
The speed of such a proton along the $x$ axis
varies strongly one from another. However, the mean velocity,
$v_{\rm pen} =\langle v_x\rangle$, of the penetrated protons remains 
more or less constant within the studied time intervals, where
$\langle v_x\rangle$ is the proton velocity which is  
averaged over proton gyroperiod. The value of
$v_{\rm {pen}}$ depends on $\alpha$ and is equal to zero (no
penetration) at $\alpha <\alpha_{\rm {cr}}\simeq 0.4$ (see Fig.\
\ref{Penetr} and Table \ref{PenetrVeloc}).

In the region $x<x_{\rm pen}$,
strong longitudinal ($E_x$) and transverse ($E_y$, $B_z$)
electromagnetic waves propagate while almost no remnants of the external
magnetic field $B_0\Theta[x]$ are found (Fig. \ref{Bfield}). Roughly, 
it could be stated that at $\alpha > \alpha_{\rm cr}$
the magnetic barrier is pushed according to the law
$B(t)=B_0\Theta[x-x_{\rm {pen}}(t)]$ and $x_{\rm {pen}}$ appears to
be a location of the barrier front at the moment $t$
as well as the particle penetration
depth. At $x>x_{\rm {pen}}$, the magnetic field remains unchanged 
except for its time-space variations due to low-frequency
electromagnetic waves which are generated by the time-variable
currents ${\bf j}$ in the front vicinity. At $\alpha\sim 1$, the 
typical amplitude of these waves is $\tilde B_0\simeq (0.2-0.3)B_0$.

In the case $\alpha<\alpha_{\rm cr}$, when the ion penetration length 
does not increase with time, all protons move
along the same track with small deviations from it. In this case, 
the velocity of all reversed protons is perpendicular to the barrier
front. However, at
$\alpha>\alpha_{\rm {cr}}$ the trajectories of protons differ
qualitatively one from another (for the angular distribution of reversed
protons in this case, see Fig. \ref{ProtAngl}). Therefore, the dimensionless 
density $\alpha$ may be called a ``stochastization parameter'' of the beam
protons. In all runs, the trajectories of electrons are very chaotic 
and differ one from another qualitatively.

\subsection{Acceleration of electrons and high-frequency radiation}

The most important feature of the dynamics of electrons 
is that they are accelerated in the barrier front vicinity 
by induced electric fields and, thus,
accumulate substantial portion of the kinetic energy of protons (see
Fig. \ref{GammaRad} and Table \ref{Results}). Figure \ref{SpectEl} shows 
the energy spectrum of outflowing electrons, $v_x<0$, in run N.
At $\alpha \sim 1$, the mean Lorentz factor of
outflowing electrons and their maximum Lorentz factor far enough
from the barrier, $x< -(c/\omega_{Bp})$, are

\begin{equation}
\langle\Gamma_e^{\rm {out}}\rangle \simeq 0.05
\left(\frac{m_p}{m_e}\right)\Gamma_0,
\quad
\Gamma_{e,{\max}}^{\rm {out}} \simeq 
{1\over 2}\left(\frac{m_p}{m_e}\right)\Gamma_0 \label{Gammas} 
\end{equation}

\noindent within a factor of 2. 
The fraction of the kinetic energy of relativistic protons
which is transformed into the energy of outflowing electrons is up to
$\sim 10\%$. The rest of the energy that is lost by protons is 
transformed into both low-frequency electromagnetic waves and 
synchrotron high-frequency radiation. 

The synchrotron high-frequency radiation is generated by single electrons
in the process of their ultrarelativistic motion in the 
electromagnetic fields. The 
mean energy of so-called radiating electrons near by the barrier front, 
$x> -(c/\omega_{Bp})$, is several times higher than the mean energy of
outflowing electrons far from the barrier, $x< -(c/\omega_{Bp})$, i.e.
$\langle \Gamma_e^{\rm rad}\rangle\simeq (3-5)\langle
\Gamma_e^{\rm out}\rangle$ (see Table II). The radiating electrons are
responsible for generation of the main part of the synchrotron
high-frequency emission. 

If the average fraction, $\xi_\gamma$, of the kinetic
energy of the plasma beam that is radiated in the vicinity of the
magnetic barrier is small, $\xi_\gamma < 0.01$ or $\Gamma_0^2B_0<3
\times 10^8$ G, the results of our simulations for $\alpha \sim 1$ 
may be fitted by the following analytic expression:

\begin{equation}
\xi_\gamma \simeq 3\times 10^{-4}(\Gamma_0/ 10^2)^2(B_0/10^3\,
\rm {G})\,.
\label{Ksi}
\end{equation}

\noindent At $\Gamma_0^2B_0 \gg 10^8$ G, the value of $\xi_\gamma$ 
tends asymptotically to $\sim 0.15$ as $\Gamma_0^2B_0$ increases. The
characteristic energy of synchrotron photons is

\begin{equation}
\langle \varepsilon _\gamma \rangle \simeq 3
({\xi_\gamma/10^{-2}})\,\,\rm {MeV} \label{Epsilon}
\end{equation}

\noindent within a factor of 2 or so. Except for numerical factors,
Eqs. (\ref{Ksi}) and (\ref{Epsilon}) can be derived
analytically from Eqs. (\ref{dExdx}) -- (\ref{DefAlpha}).

At $\alpha\ll 1$, electromagnetic fields which are induced in
the process of the beam -- barrier collision are much smaller than
$B_0$, and the beam particles may be treated as test particles
in the field of the barrier. In this case,
there is almost no energy transfer from
protons to electrons. Therefore, it is natural that both 
acceleration of electrons and their high-frequency radiation 
are strongly suppressed at $\alpha\ll 1$ as was
observed in our simulations (see Table \ref{Results}). 
Preliminary, we have the following scaling laws:
$\langle\Gamma_e^{\rm {out}}\rangle\propto\alpha$,
$\Gamma_{\rm {e},\max}^{\rm {out}}\propto\alpha$,
$\langle\Gamma_e^{\rm {rad}}\rangle\propto\alpha$, $\xi_\gamma\propto
\alpha^2$ and $\langle \varepsilon_\gamma\rangle \propto \alpha^2$.

We did not observe any decrease in both non-stationarity of the process
of the beam -- barrier interaction and in acceleration of 
electrons, even in simulations extended up to $t_{\max}\simeq 16 T_p$ 
(see Figs. \ref{ProtDens} and \ref{GammaRad}).

\section{Stationary collision}

The problem of the beam -- barrier collision is simplified significantly
if it is treated as a stationary one. This is because in a stationary 
consideration a time-dependence of all values is 
disregarded, and the beam particles of any kind (protons or electrons) move 
along exactly the same track. In this Section we consider 
the beam -- barrier interaction in a stationary manner. 

A fully relativistic self-consistent model of stationary interaction 
between a wide plasma beam and a magnetic barrier was developed by 
Peter, Ron and Rostocker in their pioneering paper
\cite{Peter79}. The set of equations
which describes such an interaction is \cite{Peter79}

\begin{equation}
\frac{dB_z}{dx}=
A_y\left(\frac{m_e/m_p}{\sqrt{z_p}}+\frac{1}{\sqrt{z_e}}\right)\,,
\label{dB_z}
\end{equation}

\begin{equation}
\frac{dE_x}{dx}=\frac{\Gamma_p}{\sqrt{z_p}}-\frac{\Gamma_e}
{\sqrt{z_e}}\,,
\label{dE_x}
\end{equation}

\begin{equation}
\frac{d\phi}{dx}=-E_x \,,\,\,\,\,\,\,\,\,\,\,
\frac{dA_y}{dx}=B_z\,,
\label{phi}
\end{equation}

\noindent
where $\Gamma_j\equiv\Gamma_0-\mu_j\phi$, $z_j\equiv\Gamma_j^2
-\mu_j^2A_y^2-1$,

\begin{equation}
\mu_j\equiv\left\{
\begin{array}{crcl}
\,\,-1\,\,\,\,\,\,\,\,\,\,\,\,\,\,\,\,\,\,\,\,\,
{\rm for\,\,electrons} \,\,(j=e)\,,\\
m_e/m_p\,\,\,\,\,\,\,\,\,\, {\rm for\,\, protons}\,\,(j=p)\,,
\end{array} \right.
\label{muj}
\end{equation}

\noindent
$\phi$ and $A_y$ are the electrostatic and magnetic potentials in units 
of $e/mc^2$. In these equations we have measured the distance $x$ in terms 
of $\sqrt{2}(\omega_{pe}\beta_0/c)$, where $\omega_{pe}=(4\pi n_0e^2
/m_e)^{1/2}$ is the electron plasma frequency and $\beta_0=v_0/c$. 

Eqs. (\ref{dB_z}) and (\ref{dE_x}) for the magnetic 
and electric fields are first order ordinary differential equations. 
Strictly speaking, such an equation allows only one boundary condition. 
For the magnetic field $B_z$, the boundary condition is $B_z=B_0$ at 
the proton turning point $x_p$ (see Fig. \ref {Entering}),
where $B_0$ is the field of the barrier. The strength of the
magnetic field $B_f$ at the front of the barrier, $x=0$, has to be 
found from integration of Eq. (\ref{dB_z}). An important issue 
is {\em where} the electric field $E_x$ should vanish: at
the right or left boundary of the collision regin $0\leq x\leq x_p$. 
In the paper \cite{Peter79}, the electric field vanishes at $x=x_p$,
$E_x|_{x=x_p}=0$. This seems quite natural and corresponds to our 
nonstationary solution. However, in Ref. \cite{Peter79} it is wrongly 
suggested that if the beam density is high enough the electric field 
$E_x$ is zero at the left boundary as well, $E_x|_{x=0}=0$. 
Below, stationary solutions for the beam -- barrier collision are 
considered in detail, and, in particular, it is shown that there is a net 
charge in the region $0\leq x \leq x_p$ irrespective 
of the beam density untill the beam -- barrier collision is stationary, 
i.e. the transverse 
electric field $E_x$ cannot be zero at the both boundaries at once.

The right-hand side of Eqs. (\ref{dB_z}) and (\ref{dE_x})
for the derivatives of the electric an magnetic fields contains terms
which tend to infinity at the electron and proton turning points, i.e. at 
both $x=x_e$ and $x=x_p$. This is the main difficulty which 
does not permit to apply a standard numerical technique to 
integration of Eqs. (\ref{dB_z}) and (\ref{dE_x}) directly. 
Since the electric and magnetic fields should remain finite everywhere,
the mentioned singularities are integrable. To integrate a set of 
equations with such a weak, integrable singularity, one should switch to
another independent variable so that all derivatives with respect 
to this new variable are finite. The interval of the beam particles may be 
used as such a new independent variable instead of $x$. In this case,
the right-hand sides of Eqs. (\ref{dB_z}) and (\ref{dE_x}) are finite 
everywhere. This is because the particle density per unitary interval 
is constant (e.g., Ref. \cite{Peter79}), rather than infinite at the
turning points if measured per unitary $x$. We integrated Eqs. (\ref{dB_z}) 
-- (\ref{phi}) from $x=0$ to $x=x_e$ using the {\em electron} interval
$s_e$ as an independent variable. Then, we switched to
another independent variable, $s_p$, which is the {\em proton}
interval measured from the {\em electron} turning point with the initial
condition $s_p|_{x=x_e}=s_e|_{x=x_e}$. We thus broke the region of
integration into two parts and used a different independent variable 
for each sub-interval to integrate Eqs. (\ref{dB_z}) -- 
(\ref{phi}). Then, we switched back to $x$ as an independent variable
to present results of this integration.

Following Ref. \cite{Peter79}, we have used the dimensionless parameter 
$\alpha_f=8\pi n_0m_pc^2\sqrt{\Gamma_0^2-1}/B_f^2$ as a measure of
the beam density instead of $\alpha$. The parameter 
$\alpha_f$ is more convenient than $\alpha$ for integration of 
Eqs. (\ref{dB_z}) -- (\ref{phi}) because such integration is performed 
from $x=0$ to $x=x_p$. In our calculations the value of $\alpha_f$
varies in a wide range from $10^{-2}$ to $10^3$. 

To integrate Eq. (\ref{dE_x}), the following
two kinds of a boundary condition for $E_x$ were specified: 
$\left.E_x\right|_{x=0}=0$ or $\left.E_x\right|_{x=x_p}=0$.
Integration of Eq. (\ref{dE_x}) for 
the boundary condition $\left.E_x\right|_{x=x_p}=0$
was carried out with a "shooting method" technique. 
The electric field was guessed at the left boundary,
$x=0$, and Eq. (\ref{dE_x}) was integrated with all the
quantities specified on that boundary. Then, the electric field on the left
boundary was adjusted to give a closer agreement with
the boundary condition $\left.E_x\right|_{x=x_p}=0$. Such shots
were repeated untill the boundary condition for the electric field was satisfied
with a desirable accuracy.  As a rule, only a few shots were required
to reach the accuracy of $\sim 1$\%.

The computational engine we used to solve the regularized set of equations
was ``LSODE'' package \cite{lsode} together with the ``OCTAVE'' 
interpreter \cite{octave}. The developed octave script is mostly 
compatible with the popular MatLab \cite{matlab} language.

Figure \ref{BEu} shows typical numerical solutions for
the longitudinal electric field $E_x$, the magnetic field $B_z$ and 
the $x$ component of the four-velocity of electrons $u_{e,x}$ in run SR10.
In this run, the boundary condition for $E_x$ is $E_x|_{x=x_p}=0$.
The energies of both electrons and protons inside of the barrier are
shown in Fig. \ref{Energype}. The results of our calculations are 
summarized in Tables IV and V. From these Tables and Fig. \ref{BEu} we can see
that the electric field $E_x$ is not equal to zero at $x=0$ and $x=x_p$
at once, as mentioned above.
A common feature of all stationary solutions is that the 
electric and magnetic fields sharply increase near the electron turning
point $x_e$ as it was pointed out in Ref. \cite{Peter79}. 
There is the following tendency: the greater the value of
$\alpha_f$ is, the sharper the rise of the fields is. Due to the
change of independent variables discussed above, the sharp rise of the fields
at the electron turning point is well-resolved: it covers a large number of
integration points and derivatives of the fields with respect to the new
independent variable remain finite (see Fig. \ref{Grid}).  The later facts 
allow us to claim a high accuracy of our calculations.
 
In the paper of Peter et al. \cite{Peter79}, Eqs. 
(\ref{dB_z}) -- (\ref{phi}) were integrated directly using $x$ as 
an independent variable, and, therefore, 
they ran into serious difficulties on their way \cite{PeterPC} which
led to certain mistakes. For example, they were not
able to calculate the value $E_x(x)$ at $x=x_e$ directly, and the
integration of Eq. (\ref{dE_x}) was treated as a two-point 
boundary value problem, adjusting $E_x(x_e)$ to be
equal to zero at the both boundaries, 
$E_x(0)=E_x(x_p)=0$. Our calculations show that such an adjustment is 
impossible (see Tables IV and V). The authors of Ref. \cite{Peter79} 
believed that at $\alpha_f\gg 1$ the
orbit of electrons is strongly elongated in the $y$ direction
in a narrow vicinity of the electron turning point, and the beam
electrons move more or less along the barrier front with 
$x\simeq x_e$ for a long time. As a result, 
a net charge in the collision region might be zero. 
We did not observe such a strong elongation of the electron orbit.
Moreover, for the boundary condition $\left.E_x\right|_{x=0}=0$
we have $\alpha_f\lesssim 1$, i.e. for this boundary condition
$\alpha_f$ cannot be much more than unit. 
 
We have confirmed the suggestion of Ref. \cite{Peter79} that 
electrons of the beam may be strongly accelerated inside the barrier. 
However, such an acceleration was observed only in runs with 
the boundary condition for $E_x$ in the form 
$\left.E_x\right|_{x=x_p}=0$. For 
the boundary condition $\left.E_x\right|_{x=0}=0$,
the beam electrons do not penetrate substantially deeper 
into the barrier than their gyroradius in the field $B_f$, and their
acceleration is very small. 

For the boundary condition $E_x|_{x=x_p}=0$, a stationary 
solution of Eqs. (\ref{dB_z}) -- (\ref{phi}) can exist 
for any nonzero value of $\alpha_f$. For all these solutions,
we have $\alpha\lesssim 0.5$ (see Fig. \ref{Alpha}). 
This upper limit on $\alpha$ is more or less evident since
in a stationary case the ram pressure of the beam cannot be more than 
the magnetic field pressure of the barrier. As to the case of 
the boundary condition $E_x|_{x=0}=0$, a stationary solution 
can exist only if $\alpha _f\lesssim 1$ and $\alpha\lesssim 0.2$. This 
reduction of the upper limit on $\alpha$ occurs
because in such a solution the electric field $E_x$ far enough from 
the barrier front is directed inside the barrier, $E_x>0$ at $x\gtrsim$
a few $\times x_e$. This field causes protons to penetrate even deeper 
into the barrier and to generate even stronger electric field. 
At $\alpha _f > 1$, at some distance from the barrier front the 
electric field inside the collision region
is stronger than the magnetic field, and the barrier cannot suppress 
penetration of protons inside itself.   

\section{Discussion}

One of the main results of our time-dependent simulations which are
presented in Sec. III is that the
outflowing electrons are strongly accelerated near the barrier front 
and accumulate substantial portion of the kinetic energy of 
protons. At $\alpha \sim 1$, when the ram pressure of the beam
is equal about the magnetic field pressure of the barrier,
the fraction of the kinetic energy of protons that is 
transferred to outgoing electrons is up to $\sim$ 10\%. 
The acceleration of outflowing electrons 
is completely due to nonstationarity of electromagnetic fields 
which are induced inside and near the barrier during the beam -- barrier
interaction. Indeed, in a stationary case the electric field is constant
in time and depends on the $x$ coordinate only, ${\bf E}= 
\{E_x(x),0,0\}$. If, as in both Ref.
\cite{Peter79} and Sec. IV, the energy losses of electrons due to
their radiation inside the barrier are ignored, the electron energy 
is a function of $x$ only, and the energy of reversed electrons
outside the barrier coincides with their initial energy before 
the beam -- barrier collision.

The bunching of reversed protons is a key element of nonstationarity
of the beam -- barrier collision. Roughly, the process of bunch formation
may be illustrated in the following way. The first protons, which are 
entered into the barrier from $t=0$ to $t\ll T_p$,   
run along almost semi-circle trajectory in almost unperturbed field 
of the magnetic barrier. For the same Lorentz-factor,
the gyroradius of electrons is much smaller than the gyroradius of
protons. Therefore,
electrons can not penetrate as deep as protons into the barrier.
The separation of electric charges induce a strong electric field $E_x$
in the $x$ direction according to Eq. (\ref{dExdx}).  
The following protons feel this electric field. The field 
$E_x$ decelerates protons and accelerates electrons.
Protons that are injected into the barrier later run along ``shorter''
trajectories and spend less time inside the barrier. As a result,
in about half of a proton gyroperiod, $t\sim {1\over 2}T_p$, most of the
protons quit the barrier almost simultaneously, and formation
of the second bunch begins, and so on.

In case of large $\alpha$, the velocities, $v_x$, of reversed protons in 
the ${x}$ direction are quite different (see Fig. \ref{ProtAngl}) so
that the bunches will decay at some distance from the barrier. Once
there is no bunching at all in low $\alpha$ case, no bunches can be
observed far from the barrier no matter what $\alpha$ is considered.

The propagation of plasma beams across a magnetic field is one of
the oldest problems in plasma physics. In spite of the long history 
of investigation (e.g., Ref. \cite{Gordeev94}), a clear model of this 
phenomenon has yet to emerge. However, some general conclusions about
the beam -- magnetic field interaction were done many years ago.
For example, if the plasma flow pressure is more than
the magnetic field pressure, it is possible for
the plasma to cross the field by purely magnetohydrodynamic principles. 
This propagation mode is equivalent to the motion of a solid conductor 
across the field. During propagation, the plasma beam picks
up and carries along the ambient plasma and magnetic field. Such models
of the beam propagation into the magnetic field are based on the 
diamagnetic properties of the plasma. Most probably, in our 
time-dependent simulations we observed such a propagation mode.
Physically, the pushing of the magnetic field is provided by strong
transverse electric currents ($j_y$) in the domain $x_{\rm {pen}}(t)
-(c/\omega_{Bp})<x<x_{\rm {pen}}(t)$, where $x_{\rm {pen}}(t)$ is the
$x$ coordinate of the most deeply penetrated particle. If $\alpha$ is 
large enough, $\alpha >\alpha_{\rm cr}$, this current layer screens
strongly the external magnetic field of the barrier at 
$x<x_{\rm {pen}}(t)-(c/\omega_{Bp})$, and moves deeper and 
deeper into the barrier. From our time-dependent simulations, 
we have $\alpha_{\rm cr}\simeq 0.4$. This is more or less consistent with
the conclusion that there are no stationary solutions for
the beam -- barrier collision if $\alpha$ is more than $\sim (0.2-0.5)$
depending on the boundary condition on the field $E_x$.

Many important features of our nonstationary solutions are absent 
in a stationary one. They are energy transfer from incoming
protons to outgoing electrons, bunching of outgoing protons, excitation 
of low-frequency electromagnetic waves, etc. 

The main question which remains open
untill now is what the final state of the beam -- barrier
system in a very long-time run, $t\gg T_p$, is. As noted,
we did not observe any decay of non-stationarity of the beam -- barrier
interaction, even in simulations extended up to $t_{\max}\simeq 16 
T_p$. However, it is possible that at $t\gg 10T_p$ our nonstationary 
solution tends to a stationary one. It is worth noting that
a stationary solution of the beam -- barrier collision might be unstable if 
$\alpha$ is high enough. Indeed, the energy of electrons inside the barrier 
may be a few hundred times higher than their initial energy. In this 
case, small ($\sim 10^{-2}$) perturbations might result in that 
some part of reversed, $v_x < 0$, electrons is stopped and captured 
inside the barrier before reaching the barrier front.  
Besides, electrons may be captured by the barrier because of 
their energy losses via synchrotron emission
in electric and magnetic fields if 
these fields are strong enough. In turn, such a capture of electrons
may result in complete destruction of the stationary solution.

The beam -- barrier
collision is similar, in many respects, to collisionless shocks. Such 
shocks in astrophysical settings can and do accelerate charged particles 
to high energies (see, for a review, Refs. \cite{be,je}). The efficiency 
of particle acceleration by shocks may be as high as $\sim 20-40$ \%
(e.g., Ref. \cite{be}). This value is more or less the same as the fraction
of the kinetic energy of the beam which may be transferred to 
high-energy electrons in the process of the beam -- barrier collision. 

Besides of analytical
calculations (e.g., Refs. \cite{Jokipii87,Ost88}),
acceleration of particles by shocks was studied numerically
in both test particle Monte Carlo simulations (see Ref. \cite{bej95} and 
references therein) and self-consistent plasma simulations (e.g., Refs.
\cite{guest88,burgess89,scholer90,kvb91}).
Most current self-consistent simulations of plasma shocks used in
astrophysics are of the hybrid type because the $m_e/m_p$ ratio
is small, $m_e/m_p\simeq 0.54\times 10^{-3}$. 
In the hybrid approach, the ions are treated
kinetically using standard particle-in-cell techniques,
while the electrons are treated as a massless,
charge neutralizing fluid. In our simulations we did not use the
hybrid approach because the case of ultrarelativistic beam is, in some
respects, easy than non-relativistic case. This is because in our case,
$\Gamma_0\gg 1$, electrons are accelerated fast, and the mean energy 
becomes only an order of magnitude smaller than the mean energy of protons.
In this case, the ratio of the gyroradii of
electrons and protons is about two orders of magnitude larger than
$m_e/m_p$ that is the ratio of the
gyroradii of electrons and protons in non-relativistic plasma.
This allowed us to treat both electrons and protons kinetically.

The configuration considered in our simulations is not quite
self-consistent. Indeed, the barrier magnetic field 
has to be generated by some kind of current
flowing in the plane $x=0$ in the $y$ direction. However, 
the evolution of this current because of its interaction with the induced
electromagnetic fields was not studied. In laboratory experiments,  
an external current flowing along a sheet of thin wires can be
the barrier-front current.
As to a situation which is relevant for astrophysics,
the barrier fields has to be generated by the magnetized plasma of the
barrier. Numerical considerations of such a configuration are under
way.

\section{Astrophysical applications}

There are at least two kinds of astrophysical objects to which the 
results of our simulations may be applied. They are the sources 
of $\gamma$-ray bursts and strongly collimated jets of active
galactic nuclei.

\subsection{Cosmological $\gamma$-ray bursters}

Gamma-ray bursts are brief, $\sim 10^{-2}-10^2$ s, bursts
of high-energy radiation that appear at random in the sky, emitting
the bulk of their energy in the range from $\sim 0.1$ MeV to a few MeV
(for a review of observational data on $\gamma$-ray bursts see Ref.
\cite{Fishman}).
During their appearance, they often outshine all other sources in
the $\gamma$-ray sky combined. More than two decades have passed since
the discovery of $\gamma$-ray bursts, but their origin is still a
mystery. The observed isotropy of $\gamma$-ray bursts in the sky and 
deficiency of faint bursts strongly suggest \cite{Briggs,arg} that they 
are cosmological in origin \cite{Usov75}.
A cosmological origin of $\gamma$-ray
bursters implies that the typical energy output in both hard X-rays and 
$\gamma$-rays is about a few $\times 10^{51}$ ergs, that is, 
about $10^{-3}M_{\odot }c^2$. Such high energetics 
of $\gamma$-ray bursters and a short time scale of $\gamma$-ray
flux variability call for very compact objects as a source of
$\gamma$-ray bursts. These objects may be either 
fast-rotating neutron stars which
arose from accretion-induced collapse of white dwarfs in binaries
\cite{Usov92} or differentially rotating disk-like objects which
are formed by the merger of a binary consisting of two neutron stars
\cite{Goodman87,Eichler89}. The strength of the magnetic field 
$B_s$ at the surface of these objects may be up to $\sim 10^{16}$ G
\cite{Usov92,TD}. 

A common point of all acceptable models of $\gamma$-ray bursters is
that a relativistic, $\Gamma_0 > 10^2-10^3$, wind is a 
source of their radiation (e.g., Refs. \cite{Hartmann,Dermer95}). 
Otherwise, high-energy photons are absorbed by $\gamma + \gamma
\rightarrow e^+ + e^-$, and both observed durations of $\gamma$-ray
bursts and their hard energy spectra, often with a very high energy
tail extending up to hundreds of MeV,
cannot be explained. 
Most probably, the radiation of $\gamma$-ray bursts is
produced in the process of interaction between such a wind and an
external medium \cite{Meszaros93,Smolsky95}.
For both the post-merger objects and fast-rotating neutron stars,
the total energy of the outflowing gas cannot be more than $10^{53}$ ergs.
Therefore, to explain the energetics of cosmological $\gamma$-ray bursters
it is needed that not less than a few per cent of the wind energy is
transformed into high-frequency radiation.

From our simulations of the beam -- magnetic barrier collision
it follows that the energy which is lost by a relativistic magnetized 
wind in the process of its interaction with an external medium 
is distributed in the following way: About half of this energy is in
ultra-relativistic protons which are reflected from the wind
front. The mean energy of reflected protons in the frame of the
burster is $\sim m_pc^2\Gamma_0^2$.  The other half of the wind
energy losses is distributed more or less evenly between low-frequency
oscillations of electromagnetic fields, and both high-energy
electrons and their high-frequency radiation. The low-frequency 
electromagnetic oscillations which are generated near the wind front
can accelerate electrons to very high energies and decay at some 
distances from the front \cite{Gunn,Kulsrud}.
As a result of these, the fraction of the
kinetic energy of relativistic magnetized winds which is transferred 
to high-energy electrons and then to high-frequency emission might
be, in principle, as high as $\sim 40 - 50\%$. This is more than enough
to explain the energetics of $\gamma$-ray bursters. 

 In the burster frame, the characteristic energy, $\langle
\tilde \varepsilon_\gamma \rangle$, of photons which are generated by
accelerated electrons in the vicinity of the wind front is $\sim
\Gamma_0$ times higher than the value of $\langle \varepsilon_\gamma
\rangle$ (see Eq. (\ref{Epsilon})) in the frame of the outflowing
gas because of the Doppler effect, $\langle \tilde \varepsilon_\gamma
\rangle\simeq \Gamma_0 \langle \varepsilon_\gamma \rangle$. Taking
into account that $\Gamma_0$ has to be more than $\sim 10^2$
\cite{Baring93,Narayan92}, from Eq. (\ref{Epsilon}) we can see that 
the radiation which is generated near the wind front with a high enough
efficiency, $\xi_\gamma \gtrsim 10^{-2}$, has the characteristic energy
$\langle\tilde \varepsilon_\gamma \rangle \gtrsim 10^2-10^3$ MeV,
while the main part of emission of detected
bursts is in the range from a few $\times 10$ keV to a few
MeV. In other words,
the radiation from the front vicinity is either too hard or too
weak to explain the bulk of the burst emission. The $\gamma$-ray
emission which is generated near the wind front may be
responsible for the high energy $\gamma$-rays, $\varepsilon_\gamma
>$ a few ten MeV, which are observed in the spectra of some bursts
\cite{Hurley94}.

The bulk of emission from $\gamma$-ray bursters may be explain by the
synchrotron radiation of high-energy electrons which
are ejected from the wind front ahead of the front.
Indeed, electromagnetic fields ahead of the wind front are the
fields of low-frequency electromagnetic waves which are generated
near the wind front due to non-stationarity of the wind --
external medium interaction. In this case, the mean value of 
the magnetic field is about $0.2 B_0$.
The mean energy of high-energy electrons reflected from the
wind front is $\sim (3-5)$ times smaller than the mean energy of
radiating electrons which are responsible for the
generation of $\gamma$-rays near the 
front. Therefore, the characteristic energy of synchrotron photons
which are generated far ahead of the wind front
is about hundred times smaller than the characteristic energy,
$\langle \tilde \varepsilon_\gamma\rangle$, of photons
from the front vicinity. This energy is of the order of 
characteristic energy of $\gamma$-ray burst emission.

\subsection{The jets of active galactic nuclei}

At radio frequencies, where VLBI can resolve the emission regions at
the milliarcsecond scale, many of radio-loud active galactic nuclei
(AGNs) exhibit compact jets (e.g., Ref. \cite{Cawthorne91}).
These jets are remarkably well collimated,
with opening angles typically less than a few degrees.
The Lorentz factor of the outflowing plasma is $3\lesssim
\Gamma_0 \lesssim 30$.

Recently, the Energetic Gamma Ray Experiment Telescope (EGRET) on the
Compton Gamma Ray Observatory (CGRO) has detected \cite{gammaAGN} 
a few ten extragalactic sources,
which are thought to be radio galaxies favorably oriented so that 
the axes of the radio jets are nearly aligned in our observing direction 
(e.g., Ref. \cite{Dermer94}). The main part of the $\gamma$-ray emission from 
the EGRET sources is in the range from a few ten MeV to a few GeV. Besides, 
very hard $\gamma$-rays at TeV energies were detected 
from Mrk 421 with the ground-based Whipple telescope \cite{Punch}. 
Hence, particles are accelerated in the jets of AGNs at least up to 
the energies of $\sim 10^9 -10^{12}$ eV.

If the jet of AGN is free (confined solely by its own inertia), it 
resembles a conic section of a spherical wind. In this case, the components 
of the magnetic field parallel and transverse to the jet velocity are
$B_\parallel \propto r^{-2}$ and $B_\perp\propto r^{-1}$, respectively,
where $r$ is the distance from the central engine which is responsible
for the AGN activity. Therefore, at large distances from the engine the 
field of the jet is mainly across its velocity, $B_\perp \ll B_\parallel$,
and our simulations may be applied to decribe the interaction between
the jet front and an external medium. As it is shown above,
in the process of such an interaction
electrons may be accelerated to high energies, $\Gamma_e\gg \Gamma_0$.
Using Eq. (\ref{Gammas}), the mean Lorentz factor of accelerated 
electrons in the AGN frame is $\langle\Gamma_e 
\rangle\sim 10^2\Gamma_0^2$. For $\Gamma_0=10$, that is the
typical value of $\Gamma_0$ for the jets of AGNs,
we have $\langle\Gamma_e \rangle\sim 10^4$.

It was suggested \cite{Dermer93,Sikora,Shaviv} that the $\gamma$-ray 
emission of jets is produced when high energy electrons 
Compton-scatter an external radiation which is either the thermal 
UV radiation of the accretion disk around a supermassive black hole 
\cite{Dermer93} or the optical and UV
emission of gas clouds surrounding the central 
engine \cite{Sikora}. In both cases the typical energy of photons which 
are scattered by the high energy electrons is  $\varepsilon_0\sim 1-10$
eV. At $\Gamma_0=10$, the mean energy of photons after scattering is
$\sim\varepsilon_0\langle\Gamma_e\rangle^2\sim 0.1-1$ GeV, 
that is consistent with the data on $\gamma$-ray 
emission from AGNs.

For $\Gamma_0\sim 30$, from Eq. (\ref{Gammas})
the maximum energy of accelerated electrons is 
$\sim m_pc^2\Gamma_0^2\simeq 10^{12}$ eV which is enough to explain
the energies of TeV $\gamma$-rays detected \cite{Punch} from Mrk 421. 

\acknowledgments
This research was supported by MINERVA Foundation, Munich / Germany.

\end{multicols}

\begin{figure}
\caption{Density of reversed protons in units of $n_0$ in run N.} 
\label{ProtDens}
\end{figure}

\begin{figure}
\caption{Energy spectrum of reversed protons at 
$x< -(c/\omega_{Bp})$ in run N.} 
\label{ProtSpect}
\end{figure}

\begin{figure}
\caption{The beam penetration depth $x_{\rm pen}$ as a function of 
time $t$ in run P (thick solid line), in run N (thin solid line),
in run L (dotted line).}
\label{Penetr}
\end{figure}

\begin{figure}
\caption{Distribution of magnetic field in run N at the moments
$t=0$ (dotted line), $t=7.96 T_p$ (thin solid line) and $t=15.9 T_p$
(thick solid line).} 
\label{Bfield}
\end{figure}

\begin{figure}
\caption{Angular distribution of reversed protons 
at $x< -(c/\omega_{Bp})$ in run N.}
\label{ProtAngl}
\end{figure}

\begin{figure}
\caption{Maximum energy of accelerated electrons (thick line)
and intensity of their synchrotron radiation per unitary area of the
barrier front (thin line) in run N.}
\label{GammaRad}
\end{figure}

\begin{figure}
\caption{Energy spectrum of outflowing electrons, $v_x<0$, in run N
at the moment $t=15.9T_p$.}
\label{SpectEl}
\end{figure}

\begin{figure}
\caption{Schematic drawing of a plasma beam entering a magnetic
barrier.}
\label{Entering}
\end{figure}

\begin{figure}
\caption{Plots of the magnetic field $B_z(x)$ (thin solid line),
the longitudinal electric field $E_x(x)$ (dotted line), and
the $x$ component of the four-velocity of electrons $u_{e,x}(x)$
(thick solid line) in run SR10.}
\label{BEu}
\end{figure}

\begin{figure}
\caption{Plots of the proton energy (solid line) and the
electron energy (dotted line) in run SR10.}
\label{Energype}
\end{figure}

\begin{figure}
\caption{Numerical resolution of the main results of the stationary case, 
run SR10. $\circ$ -- magnetic field profile, $\bullet$ -- electric
field profile, $\times$ -- longitudal component four-velocity of
electrons. Each figure stands for five integration points.}
\label{Grid}
\end{figure}

\begin{figure}
\caption{The value of $\alpha$ as a function of $\alpha_f$ for
both the boundary condition $E_x|_{x=x_p}=0$ (thin line) and
for the boundary condition $E_x|_{x=0}=0$ (thick line).}
\label{Alpha}
\end{figure}

\begin{table}
\caption{Input parameters of simulations}
\label{Params}
$$
\begin{array}{ccddccd}
\rm{Run} &   B_0   &\Gamma_0&\alpha& -x_{\min}&x_{\max}&t_{\max}\\
         &(\rm {G})&(10^3)  &&(c/\omega_{Bp})&(c/\omega_{Bp})&(T_p)\\
\tableline
\rm {A} &   100   &   0.1  &   2  &    5     &   10  &    2.8  \\
\rm {B} &   300   &   0.3  &   2  &    5     &   10  &    2.8  \\
\rm {C} &   300   &   1    &   2  &    5     &   12  &    3.3  \\
\rm {D} &   10^4  &   0.5  &   2  &    5     &   7   &    3.3  \\
\rm {E} &   10^4  &   1    &   2  &    10    &   7   &    3.5  \\
\rm {F} &   10^4  &   2    &   2  &    5     &   5   &    2.4  \\
\rm {G} &   10^5  &   1    &   2  &    10    &   5   &    2.3  \\
\rm {H} &   10^4  &   1    &   1  &    10    &   5   &    4.4  \\
\rm {I} &   10^3  &   1    &   4  &    5     &   20  &    3.5  \\
\rm {J} &   10^4  &   1    &  0.2 &    5     &   15  &    5.1  \\
\rm {K} &   300   &  0.3   &  1/3 &    10    &   5   &    10.6 \\
\rm {L} &   300   &  0.3   &  2/5 &    10    &   5   &    10.6 \\
\rm {M} &   300   &  0.3   &  4/7 &    5     &   30  &    10.6 \\
\rm {N} &   300   &  0.3   &  2/3 &    5     &   30  &    15.9 \\
\rm {O} &   300   &  0.3   &   1  &    5     &   30  &    10.6 \\
\rm {P} &   300   &  0.3   &   2  &    5     &   30  &    10.6 \\

\end{array}
$$
\end{table}

\begin{table}
\caption{Derived parameters of simulations}
\label{Results}
$$
\begin{array}{ccccddd}
\rm{Run} &{\langle\Gamma^{\rm {out}}_e\rangle\over\Gamma_0}&
\Gamma_{\rm {e},\max}^{\rm {out}}\over\Gamma_0&
\langle\Gamma_e^{\rm {rad}}\rangle\over\Gamma_0&
\langle\Gamma_p^{\rm {out}}\rangle\over\Gamma_0&
\xi_\gamma & \langle\varepsilon_\gamma \rangle   \\
& & & & & (10^{-2}) & (\rm {MeV}) \\
\tableline
\rm {A}& 132 & 603 & 643 & 0.61 &3.2\times 10^{-3}& 0.013\\
\rm {B}& 107 & 785 & 809 & 0.41 & 0.12            & 0.45 \\
\rm {C}& 106 & 464 & 543 & 0.57 & 0.96            & 5.1  \\
\rm {D}& 119 & 745 & 425 & 0.48 & 3.4             & 8.9  \\
\rm {E}& 92  & 542 & 337 & 0.48 &  11             & 14   \\
\rm {F}& 92  & 785 & 235 & 0.40 &  12             & 35   \\
\rm {G}& 56  & 402 & 193 & 0.43 &  14             & 36   \\
\rm {H}& 116 & 801 & 419 & 0.57 & 9.9             & 15   \\
\rm {I}& 105 & 427 & 555 & 0.41 & 2.9             & 9.1  \\
\rm {J}& 9.5 & 39  & 41  & 0.99 & 0.16            & 0.20 \\
\rm {K}& 115 & 573 & 288 & 0.94 & 0.017           & 0.09 \\
\rm {L}& 424 & 1473& 596 & 0.73 & 0.052           & 0.19 \\
\rm {M}& 296 & 1038& 913 & 0.69 & 0.18            & 0.64 \\
\rm {N}& 489 & 1275& 807 & 0.50 & 0.36            & 0.34 \\
\rm {O}& 213 & 1284& 821 & 0.62 & 0.16            & 0.34 \\
\rm {P}& 94  & 550 & 883 & 0.63 & 0.42            & 0.68 \\
\end{array}
$$
\end{table}

\begin{table}
\caption{Penetration of the beam particles with $\Gamma_0=300$ into
the magnetic barrier with $B_0=300$ G for different values of $\alpha$.}
\label{PenetrVeloc}
$$
\begin{array}{ccccccccc}
    \alpha      &   2  & 1 & 2/3 &  4/9 & 2/5 & 1/3 & \\
\tableline
v_{\rm {pen}}/c& 0.32 & 0.17&0.077& 0.05& 0& 0 \\
\end{array}
$$
\end{table}

\begin{table}
\caption{Parameters of simulations of stationary collision
between the beam with $\Gamma_0=300$ and the magnetic barrier for the
boundary condition $E_x|_{x=x_p}=0$.}
\label{ResultsStat}
$$
\begin{array}{ccccccccc}
\rm {Run} & \alpha_f & \alpha & {\Gamma_{p,min}\over \Gamma_0} 
& {\Gamma_{e,max}\over \Gamma_0} &
{B_f\over B_0} & {E_f\over B_0} & {E_{max}\over B_0} \\
\tableline
\rm {SR1} & 0.01 & 0.0098 & 0.99 & 1.02& 0.99  & -0.015& -0.015 \\
\rm {SR2} & 1/10 & 0.086  & 0.92 & 1.2 & 0.92  & -0.13 & -0.13 \\
\rm {SR3} & 1/5  & 0.15   & 0.86 & 1.3 & 0.87  & -0.22 & -0.22 \\
\rm {SR4} & 1/2  & 0.29   & 0.76 & 2.1 & 0.76  & -0.40 & -0.41 \\
\rm {SR5} & 1    & 0.44   & 0.67 & 7.1 & 0.67  & -0.57 & -0.58 \\
\rm {SR6} & 2    & 0.50   & 0.64 & 48  & 0.50  & -0.50 & -0.62 \\
\rm {SR7} & 4    & 0.51   & 0.64 & 97  & 0.36  & -0.38 & -0.61 \\
\rm {SR8} & 10   & 0.51   & 0.63 & 170 & 0.23  & -0.27 & -0.58 \\
\rm {SR9} & 40   & 0.51   & 0.63 & 276 & 0.11  & -0.20 & -0.53 \\
\rm {SR10}& 1000 & 0.51   & 0.62 & 434 & 0.023 & -0.14 & -0.45

\end{array}
$$
\end{table}

\begin{table}
\caption{Parameters of simulations of stationary collision
between the beam with $\Gamma_0=300$ and the magnetic barrier for the
boundary condition $E_x|_{x=0}=0$.}
\label{ResultsStatLeft}
$$
\begin{array}{ccccccccc}
\rm {Run} & \alpha_f & \alpha & {\Gamma_{p,max}\over \Gamma_0} &
{B_f\over B_0} & {E_x(x_p)\over B_0} \\
\tableline
\rm {SL1} & 0.01   & 9.8 \times 10^{-3}   & 1    & 0.99 & 0.016 \\
\rm {SL2} & 1/10   & 0.082                & 1.06 & 0.90 & 0.15  \\
\rm {SL3} & 1/5    & 0.13                 & 1.1  & 0.81 & 0.29  \\
\rm {SL4} & 1/3    & 0.16                 & 1.3  & 0.69 & 0.45  \\
\rm {SL5} & 1/2    & 0.15                 & 1.5  & 0.55 & 0.63  \\
\rm {SL6} & 2/3    & 0.11                 & 1.9  & 0.41 & 0.78  \\
\rm {SL7} & 3/4    & 0.085                & 2.2  & 0.34 & 0.85  \\
\rm {SL8} & 0.8    & 0.068                & 2.5  & 0.29 & 0.88  \\
\rm {SL9} & 0.9    & 0.032                & 3.7  & 0.19 & 0.95  \\
\rm {SL10}& 1      & 4.6 \times 10^{-4}   & 31   & 0.02 & 1

\end{array}
$$
\end{table}

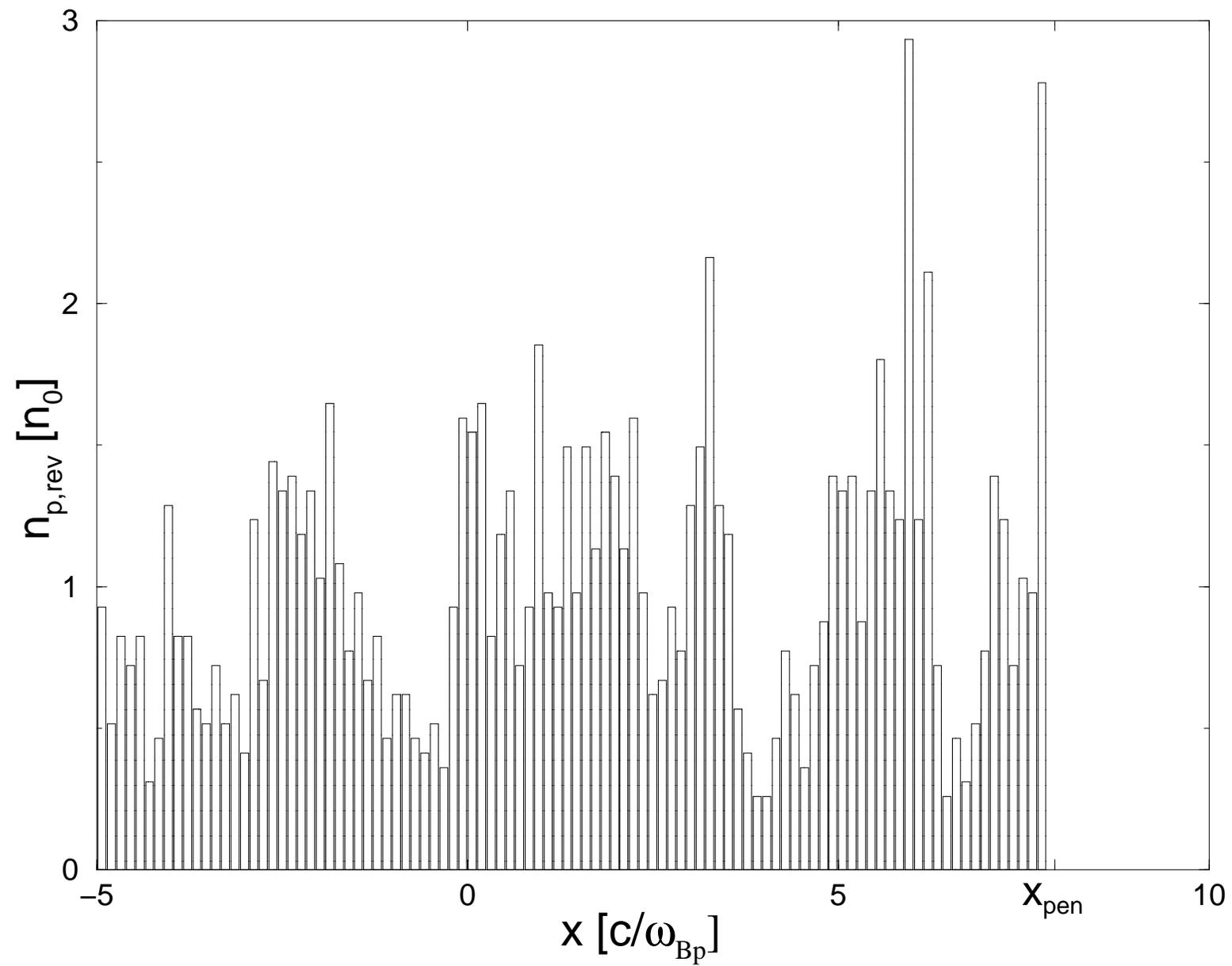
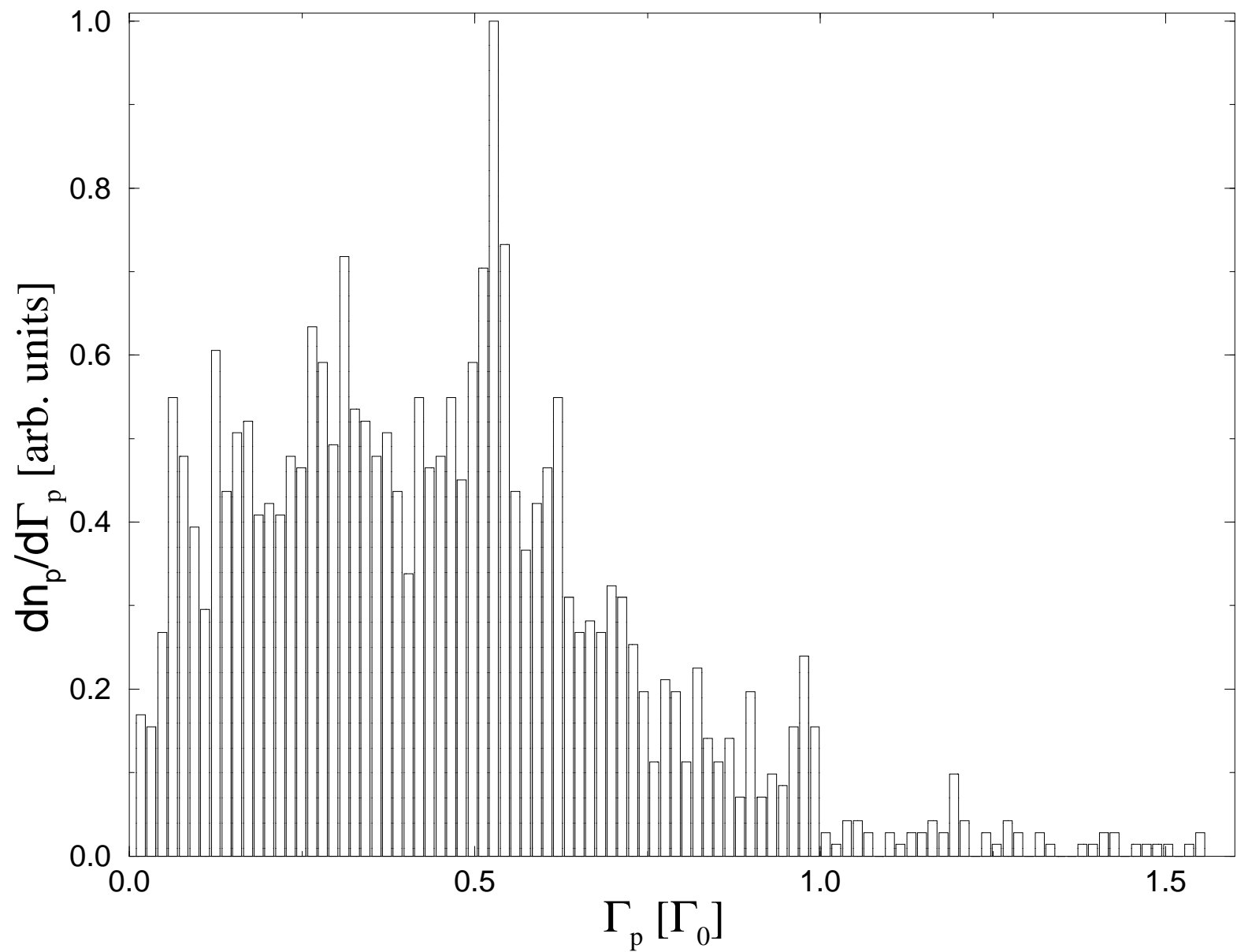
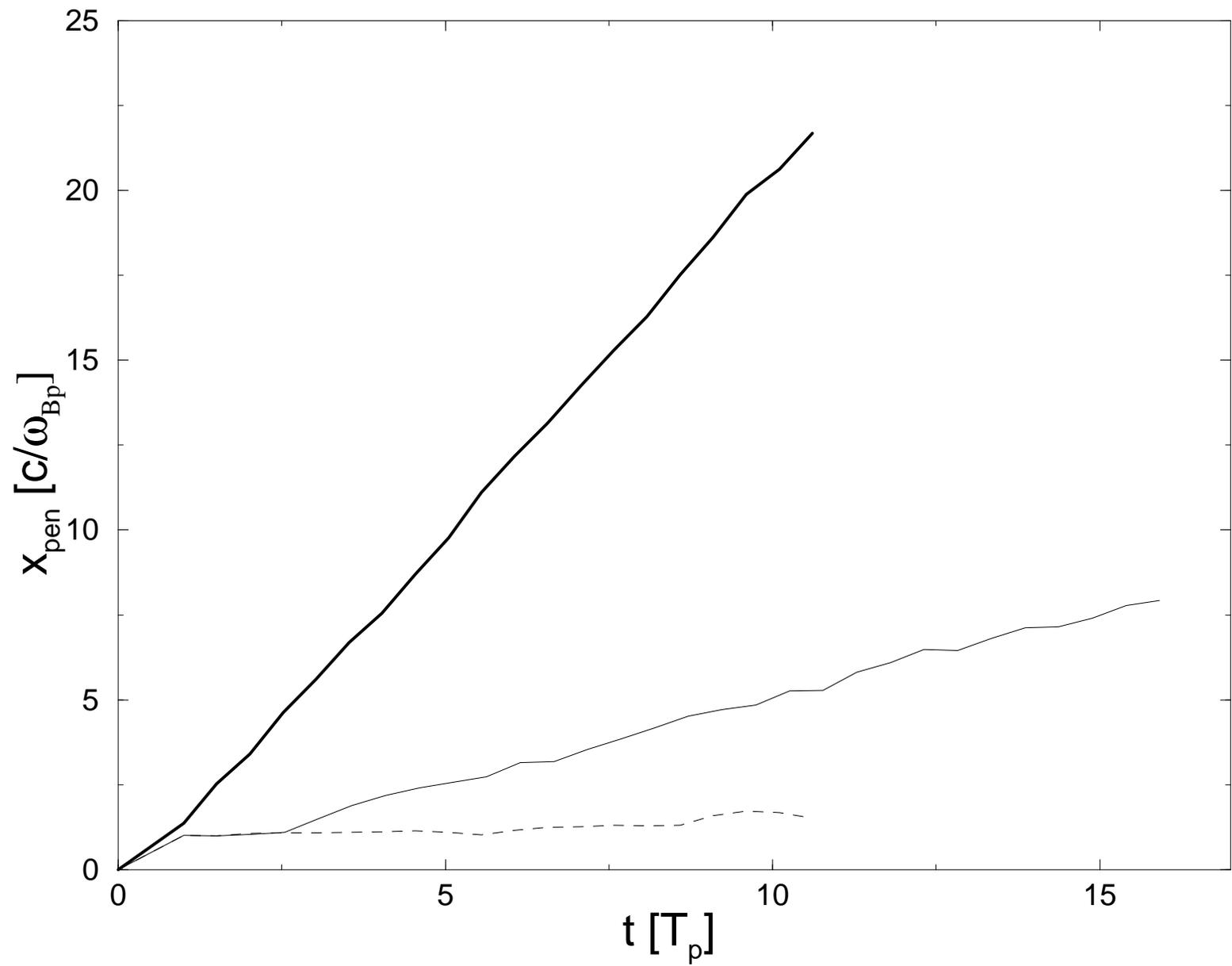
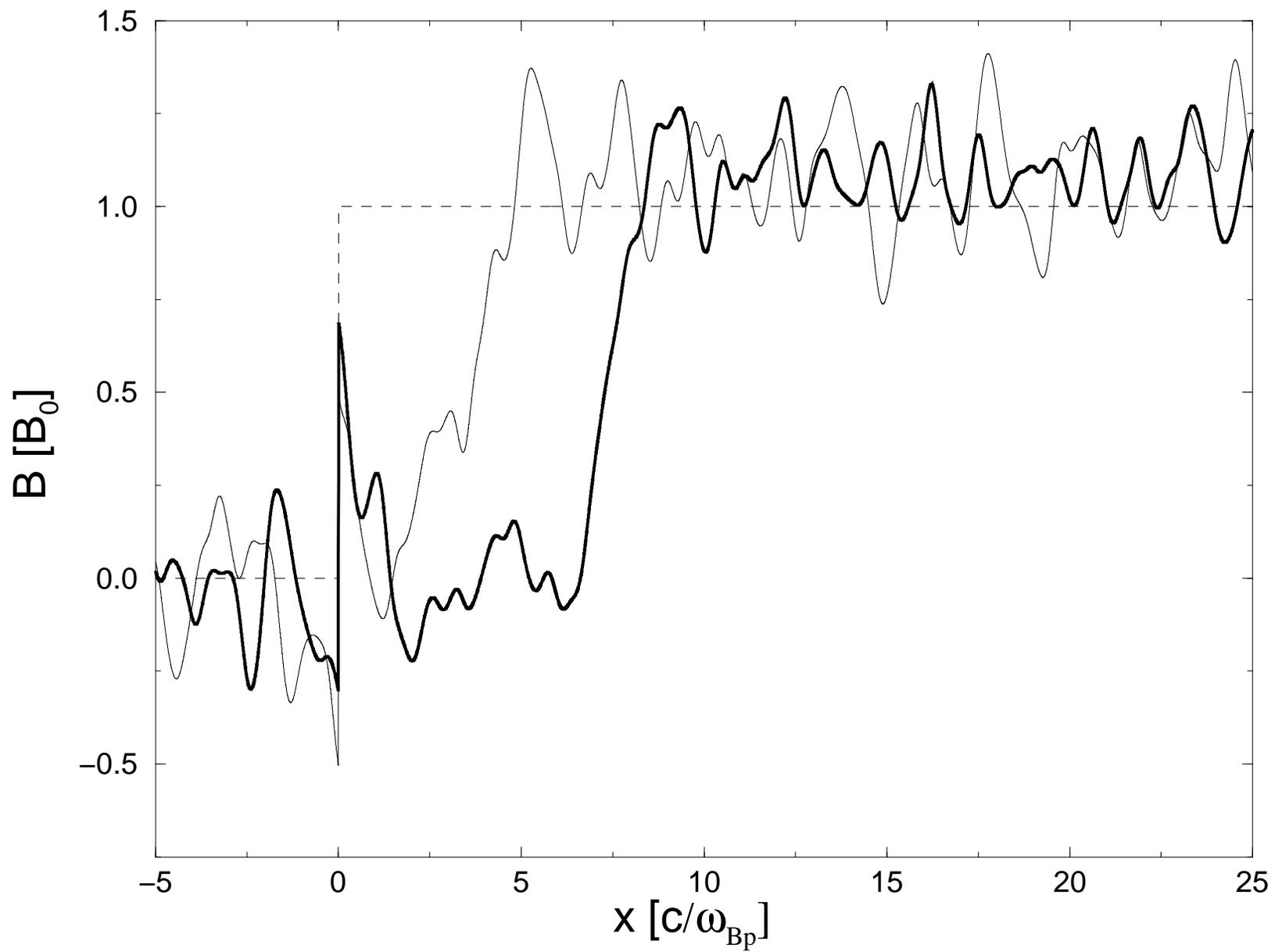
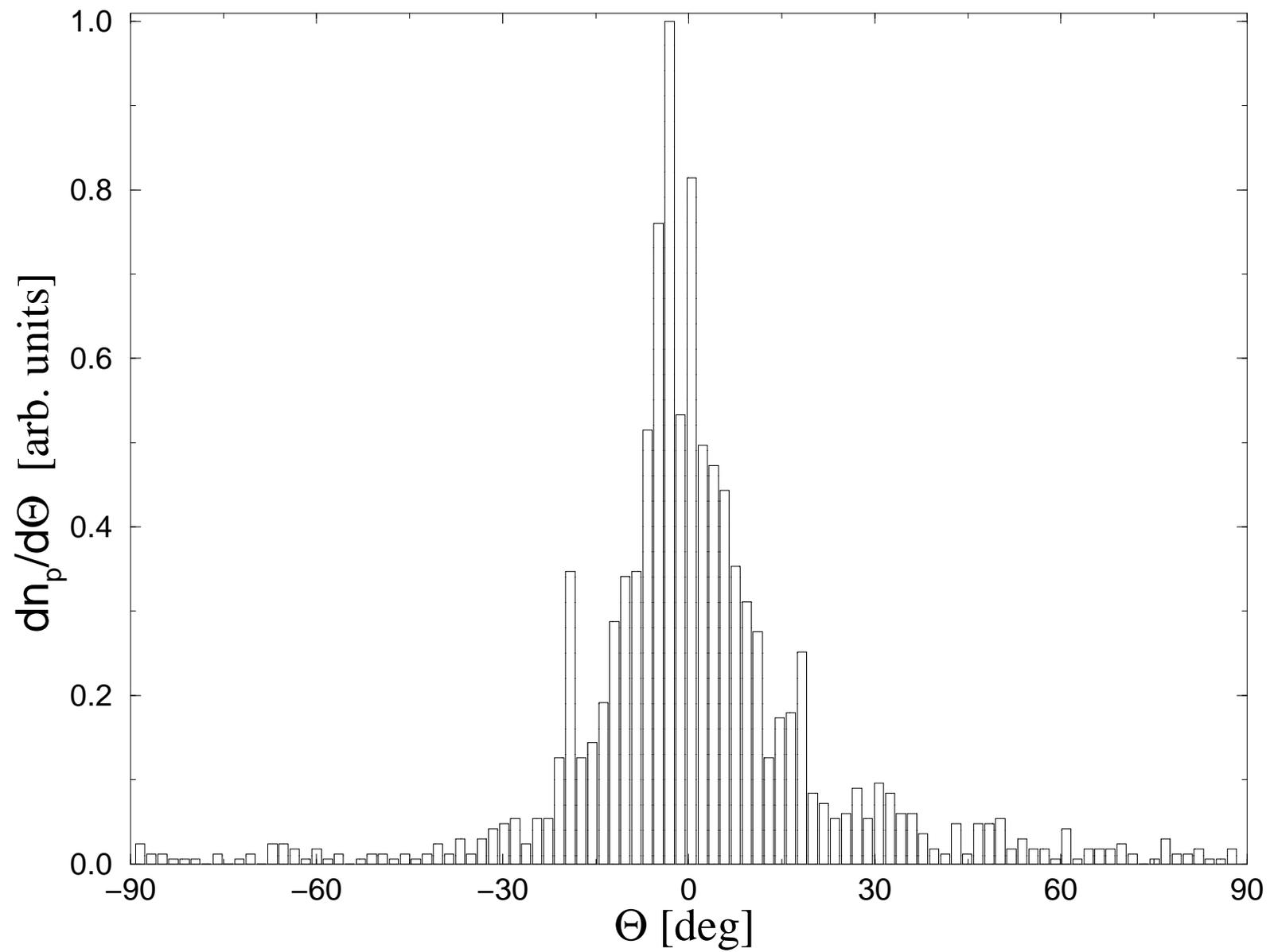
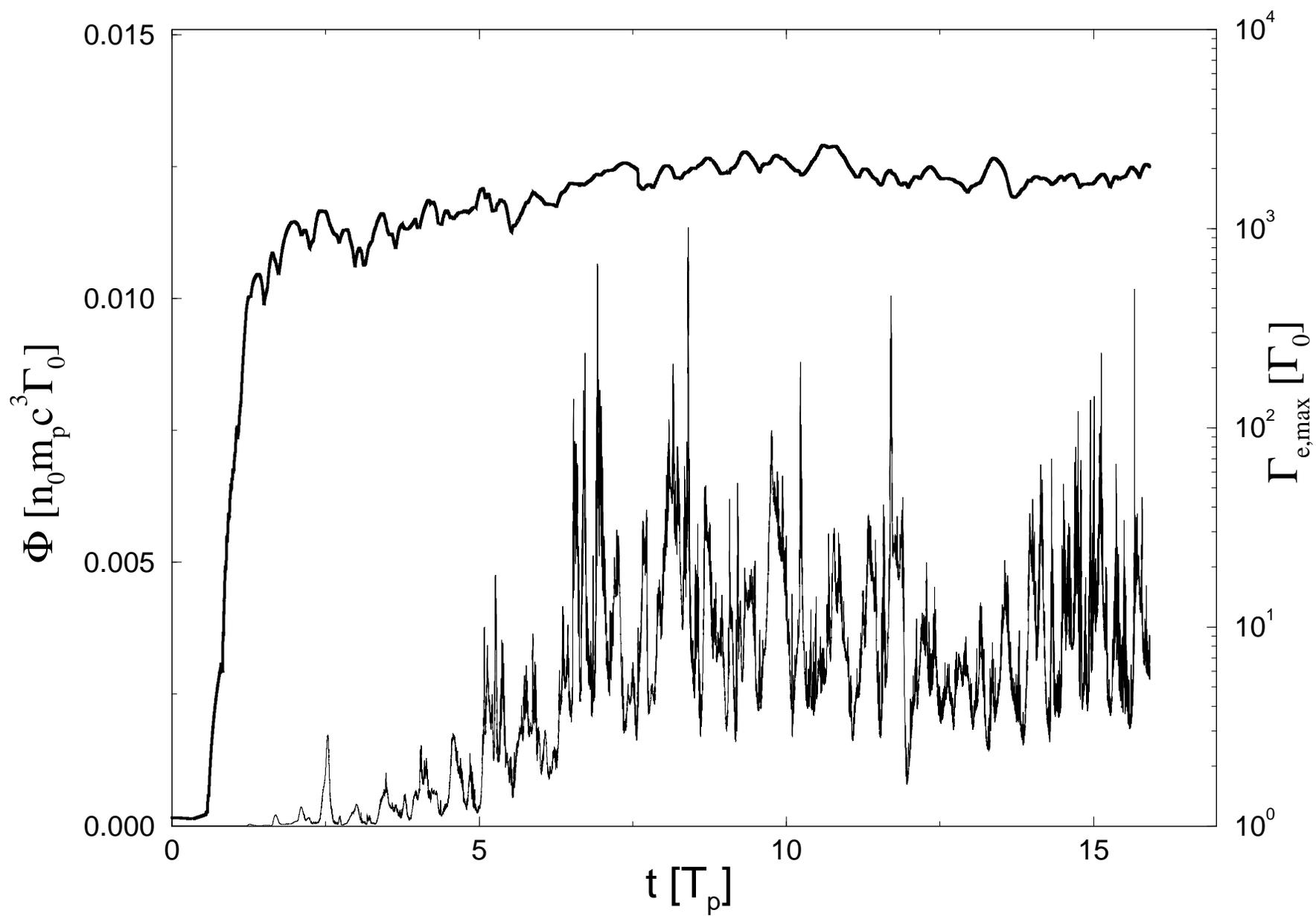
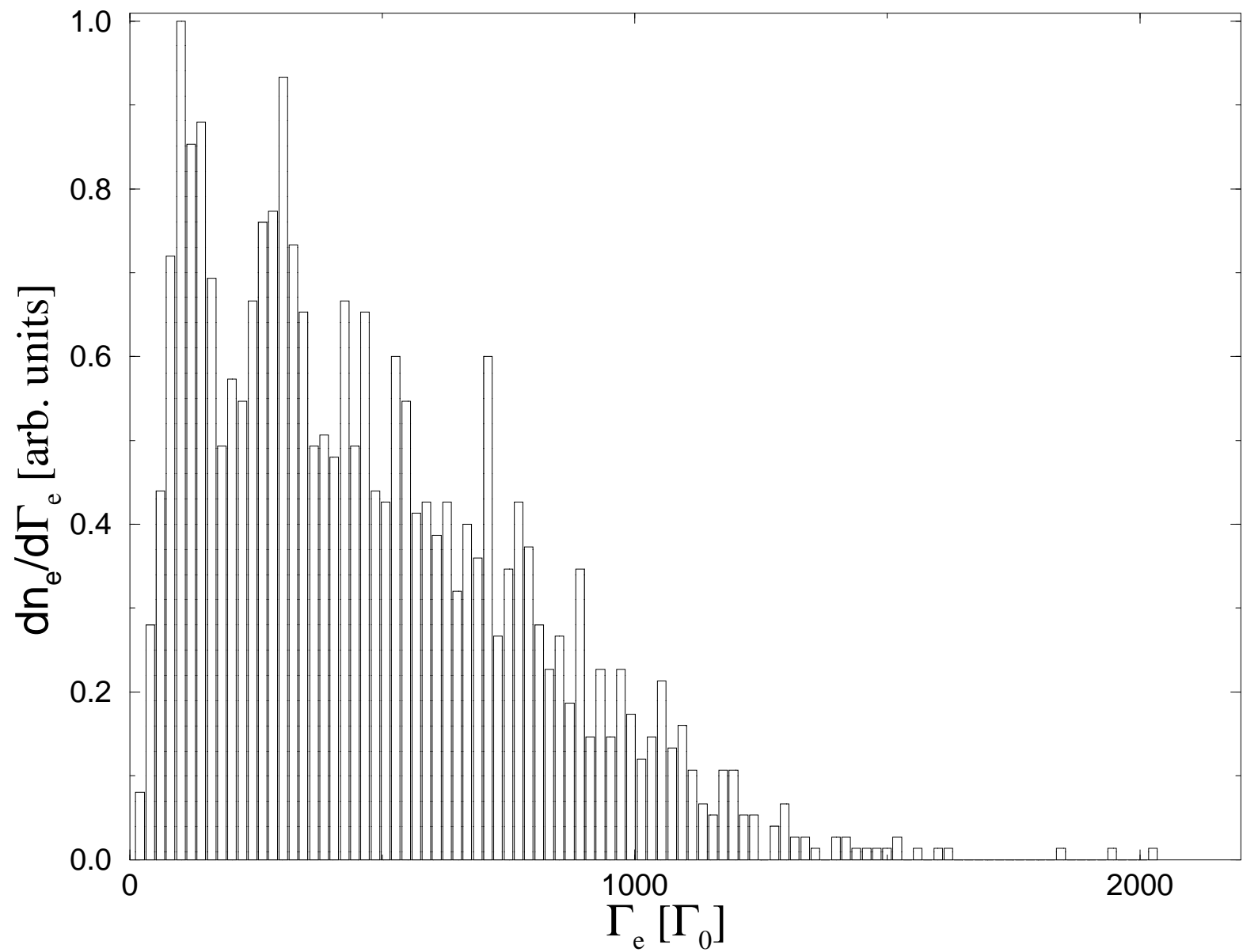
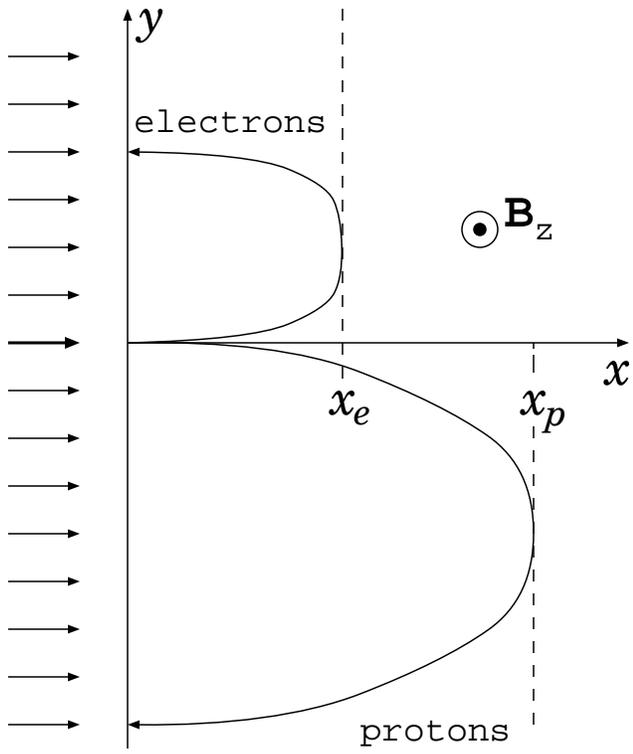
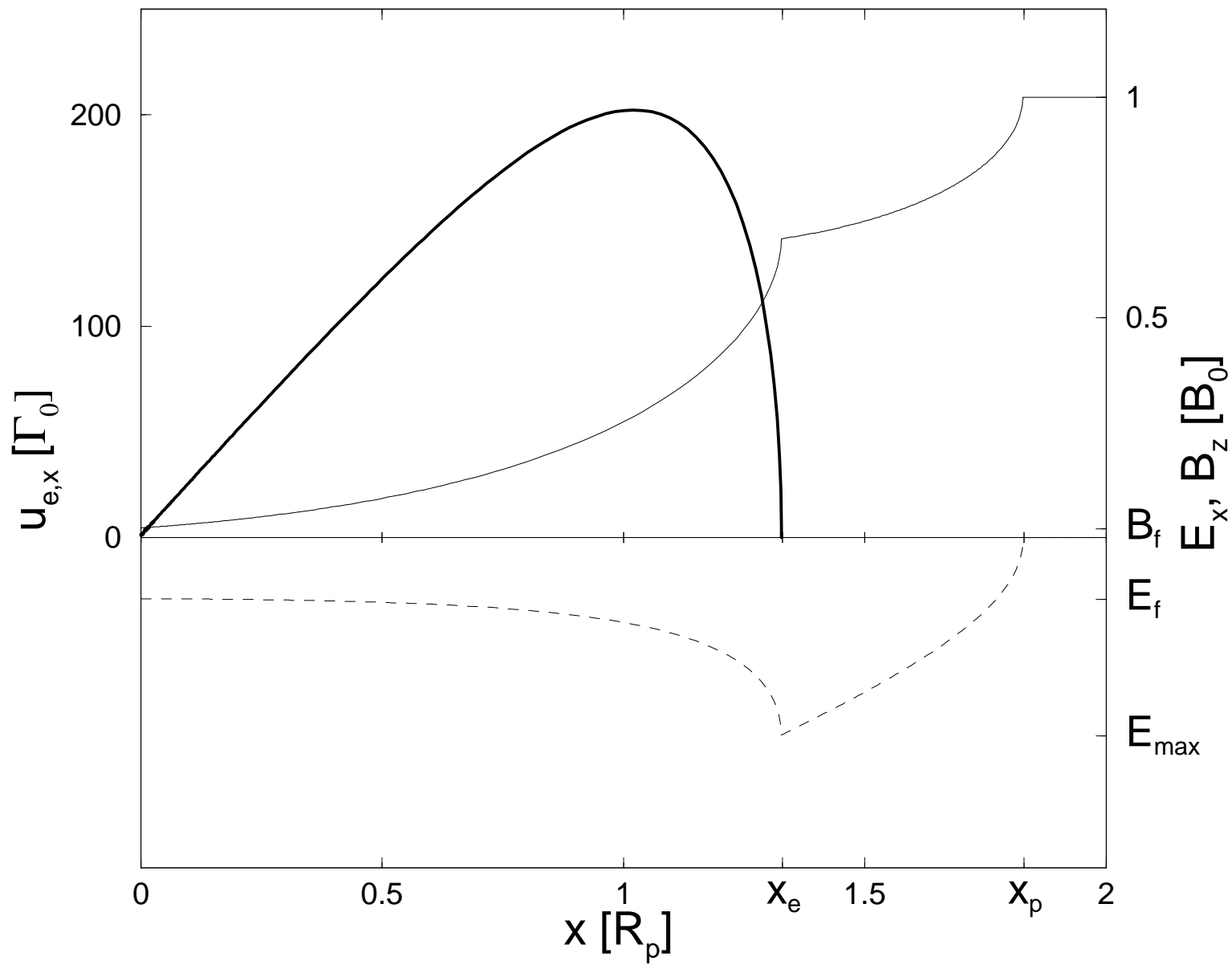
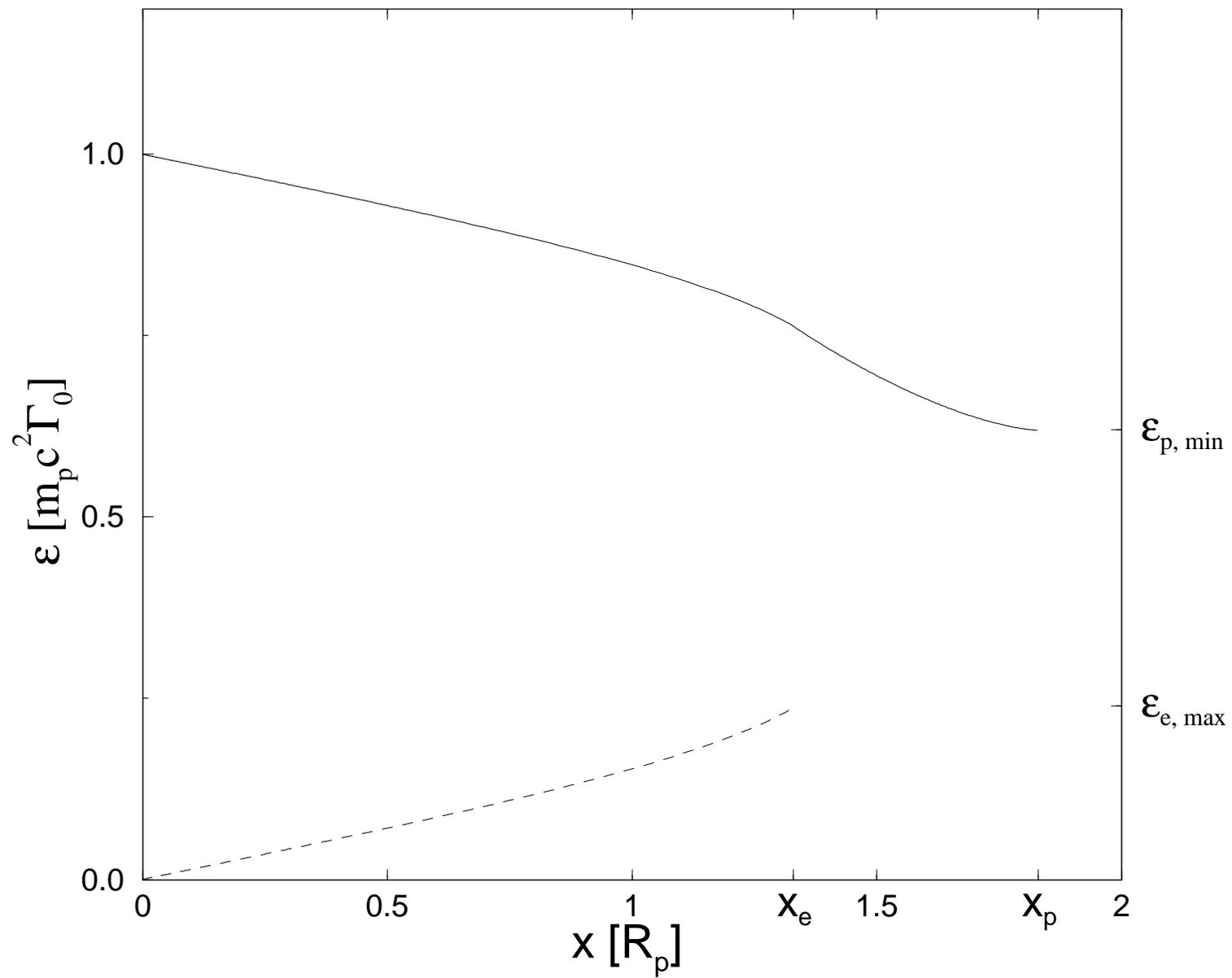
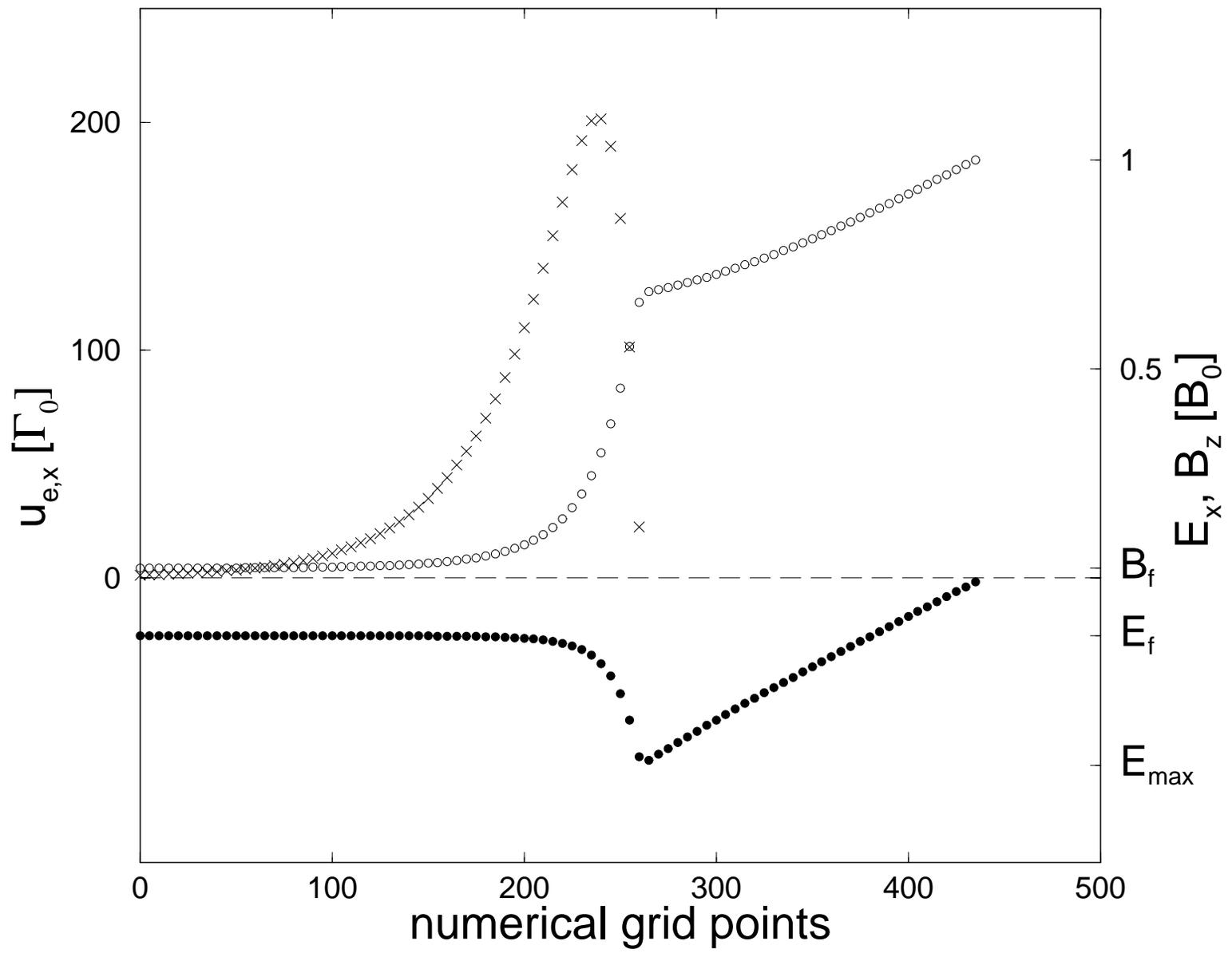
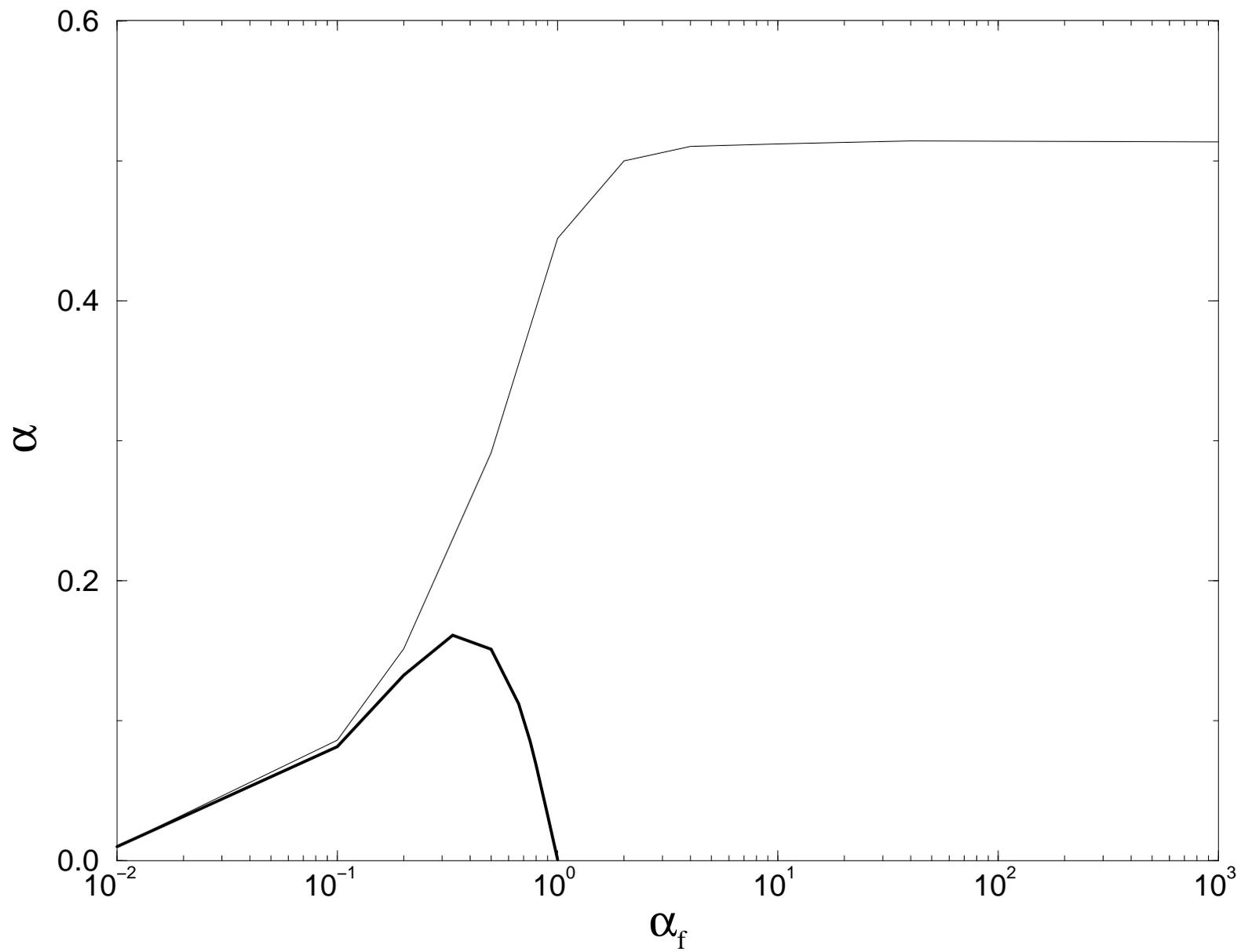
\end{document}